\documentclass[traditabstract,rnote]{aa}

\usepackage{amssymb}
\usepackage{graphicx}
\usepackage{amsmath}
\usepackage{natbib}
\usepackage{txfonts}
\usepackage{fixltx2e}
\usepackage{color}

\begin{document}

\def\na{New A, }%
\def\apj{ApJ, }%
\def\aap{A\&A, }%
\def\aapr{A\&ARv, }%
\def\apjs{ApJS, }%
\def\apjl{ApJ, }%
\def\aj{AJ, }%
\def\pasp{PASP, }%
\def\ssr{Space~Sci.~Rev., }%
\def\prd{Phys.~Rev.~D, }%
\def\mnras{MNRAS, }%
\def\memsai{Mem.~Soc.~Astron.~Italiana, }%
\def\araa{ARA\&A, }%
\def\nat{Nature, }

\def\pasj{PASJ, }

\def\psrb{PSR B1259-63/LS 2883}
\def\source{1ES1440+122}
\def\flux{\textrm{TeV}^{-1}\textrm{cm}^{-2}\textrm{s}^{-1}}
\def\gammaray{$\gamma$-ray}
\def\gammarays{$\gamma$-rays}
\def\gaga{$\gamma\gamma$}
\def\hess{H.E.S.S.}
\def\chandra{\emph{Chandra}}
\def\xmm{\emph{XMM-Newton}}
\def\asca{\emph{ASCA}}
\def\fermi{\emph{Fermi}-LAT}
\def\deg{$^{\circ}$}

\newcommand{\IS}[1]{\textcolor{red}{\textbf{\boldmath#1\unboldmath}}}

\title{Probing cluster environments of blazars through gamma-gamma absorption}

\author{Iurii Sushch \inst{1,2} 
\and Markus B\"ottcher \inst{1,3}}
\institute{Centre for Space Research, North-West University, Potchefstroom 2520, South Africa
\and
Astronomical Observatory of Ivan Franko National University of L'viv, vul. Kyryla i Methodia, 8, L'viv 79005, Ukraine
\and
Astrophysical Institute, Department of Physics and Astronomy, Ohio University, Athens, OH 45701, USA
}
\date{Received 24 September 2014/ Accepted 12 October 2014}
\abstract
{Most blazars are known to be hosted in giant elliptic galaxies, but their cluster environments 
have not been thoroughly investigated. Cluster environments may contain radiation fields of low-energy 
photons created by nearby galaxies and/or stars in the intracluster medium that produce 
diffuse intracluster light. These radiation fields may absorb very high energy $\gamma$\,rays 
($E\gtrsim100$\,GeV; VHE) and trigger pair cascades with 
further production of subsequent generations of 
$\gamma$\,rays with lower energies via inverse Compton scattering on surrounding radiation fields 
leaving a characteristic imprint in the observed spectral shape. The change of the spectral shape 
of the blazar reflects the properties of its ambient medium. We show, however, that neither 
intracluster light nor the radiation field of an individual nearby galaxy can cause substantial 
\gaga\,absorption. Substantial \gaga\,absorption is possible only in the case of multiple, 
$\gtrsim5$, luminous nearby galaxies. This situation is not found in the local Universe, but may be 
possible at larger redshifts ($z\gtrsim2$). Since VHE $\gamma$\,rays from such distances are expected 
to be strongly absorbed by the extragalactic background light, we consider possible signatures of 
\gammaray\ induced pair cascades by calculating the expected GeV flux which appears to be below 
the \fermi\ sensitivity even for $\sim10$ nearby galaxies.}

\keywords{radiation mechanisms: non-thermal -- BL Lacertae objects: general -- BL Lacertae objects: individual: \source\ -- galaxies: clusters: general}





\authorrunning{I. Sushch \& M. B\"ottcher}
\titlerunning{Gamma-gamma absorption in galaxy clusters}
\maketitle


\section{Introduction}
\label{intro}
Blazars are the brightest objects among active galactic nuclei (AGN) because of a relativistic
jet pointed close to the line of sight. They are characterized by highly variable, non-thermal 
dominated emission across the entire electromagnetic spectrum, from radio through $\gamma$-ray 
frequencies. They dominate the extragalactic $\gamma$-ray sky both at GeV energies, observed 
by the {\it Fermi} Large Area Telescope (LAT), and in very-high-energy (VHE: $E > 100$~GeV) 
$\gamma$\,rays, observable by ground-based Cherenkov telescope facilities. 

VHE $\gamma$\,rays from sources at cosmological distances are subject to absorption by 
$\gamma\gamma$ pair production on various target photon fields \citep{1967PhRv..155.1404G},
with photons of energy $E = 1 \, E_{\rm TeV}$~TeV primarily interacting with target photons
of wavelength $\lambda_{\rm target} = 2.4 \, E_{\rm TeV} \, \mu$m. Hence, photons in the
energy range $\sim 0.1$ -- 1~TeV interact predominantly with optical -- near-infrared light. 
The extragalactic background light (EBL) is usually considered to be the dominant target 
photon field for $\gamma\gamma$ absorption of cosmological VHE $\gamma$-ray sources
\citep[e.g.][]{1992ApJ...390L..49S,1999APh....11...93P,2008A&A...487..837F,2010ApJ...712..238F},
and this effect is commonly expected to limit the VHE $\gamma$-ray horizon, out to which VHE
$\gamma$-ray sources are expected to be observable with ground-based Cherenkov telescope 
facilities, to $z \lesssim 0.5$. However, the EBL is not expected to be perfectly homogeneous
and, in particular, galaxies near the line of sight to the blazar, or the cluster environment
of the blazar, may provide additional target photon fields which may potentially increase the 
expected $\gamma\gamma$ opacity. In this paper, we study the $\gamma\gamma$ opacity provided
by intervening, individual galaxies, as well as the collective photon fields provided by the
cluster environment of the blazar's host galaxy. 

Most blazars are known to be hosted in giant Elliptical galaxies, but their cluster environments 
are poorly characterized. They might contain low-energy (optical, infrared(IR)) radiation fields 
created e.g. by a nearby galaxy or stars which escaped their galaxies 
\citep[diffuse intracluster light, ICL; e.g.][]{2012MNRAS.425.2058B}, 
which may absorb VHE $\gamma$\,rays leaving a characteristic imprint in the observed spectrum. Moreover, 
such radiation fields can trigger pair cascades as the electron-positron pairs produced in 
\gaga\,absorption may generate further generations of $\gamma$\,rays via inverse Compton (IC) 
scattering on the surrounding radiation fields. In the case of effective pair cascade development, 
a considerable fraction of the energy in VHE photons is re-emitted at lower energies, significantly 
changing the spectral shape of the blazar emission. The change of the spectral shape reflects the 
properties of the ambient medium, in particular the properties of the radiation field and the 
magnetic field \bibpunct[ ]{(}{)}{,}{a}{}{,}\citep[see e.g.][for a discussion of $\gamma\gamma$ absorption and cascading
within the immediate AGN environment]{2010ApJ...717..468R,2011ApJ...728..134R}.


In Section \ref{ICL}, we discuss the effect of the ICL, while Section \ref{galaxies} contains
the discussion of the effect of individual galaxies, applying our considerations to the specific
example of the VHE $\gamma$-ray blazar 1ES 1440+122 in Section \ref{1440}. We briefly discuss
the effect of $\gamma\gamma$ induced pair cascades in cluster environments in Section \ref{cascades},
and summarize our findings in Section \ref{summary}. 
The following cosmological parameters are used throughout the paper: Hubble constant, 
$H_{0} = 67.3$\,km\,s$^{-1}$\,Mpc$^{-1}$, mean mass density, $\Omega_{\mathrm{m}} = 0.315$, and 
dark energy density, $\Omega_\lambda = 0.686$ \citep{planck}.

\section{\label{ICL}Intracluster light}

In this section we consider the possible \gammaray\ absorption due to the 
ICL. To calculate the optical depth due to \gaga\,absorption we define the 
region of the ICL as a sphere with radius $R_{\mathrm{ICL}}$ with an isotropic 
distribution of photons and a constant photon density. For a photon spectrum 
peaking at an effective energy $\epsilon_{\mathrm{eff}} = E_{\mathrm{eff}}/(m_{\mathrm{e}}c^2)$ 
the differential photon density can be approximated by
\begin{equation}
\label{density}
n(\epsilon, \mathbf{r}) = \frac{L_{\mathrm{ICL}}}{\pi R_{\mathrm{ICL}}^2 c 
\epsilon_{\mathrm{eff}}^2 m_{\mathrm{e}}c^2} H(R_{\mathrm{ICL}} - r) 
\delta(\epsilon - \epsilon_{\mathrm{eff}}),
\end{equation} 
where $L_{\mathrm{ICL}}$ is the total luminosity of the ICL in the region considered, 
$r$ is the distance from the centre of the ICL region, $m_\mathrm{e}$ is the electron 
mass, $c$ is the speed of light, and $H(x)$ is the Heaviside function ($H(x) = 1$ if $x>0$ and 
$H(x) = 0$ otherwise). For a rough estimate of the optical depth we use a delta-function 
approximation of the \gaga\,cross section of two photons with normalized energies $\epsilon_1$ 
and $\epsilon_2$ (where $\epsilon = h\nu/(m_{\mathrm{e}}c^2)$),
\begin{equation}
\label{cross}
\sigma_{\gamma\gamma}(\epsilon_1, \epsilon_2) = \frac{1}{3} \sigma_{\mathrm{T}} \epsilon_1 
\delta\left(\epsilon_1 - \frac{2}{\epsilon_2}\right),
\end{equation}
where $\sigma_{\mathrm{T}}$ is the Thompson cross section. The \gaga\ optical depth for a 
\gammaray\ photon with normalized energy $\epsilon_{\gamma}$ in a photon field with a 
differential photon density $n(\epsilon, \Omega, x)$ is given by \citep{1967PhRv..155.1404G}
\begin{equation}
\tau_{\gamma\gamma} = \int dx \int d\Omega(1-\mu)\int d\epsilon\,n(\epsilon,\Omega,x) \, 
\sigma(\epsilon, \epsilon_{\gamma}, \mu),
\end{equation}
where $dx$ is the differential path travelled by \gammaray\ photon, $d\Omega$ is the solid 
angle element, $\mu = \cos\theta$, and $\theta$ is the interaction angle between the \gammaray\ 
photon and a target photon. Taking into account Equations (\ref{density}) and (\ref{cross}) 
and assuming that the \gammaray\ source is located in the centre of the ICL region, the \gaga\ 
optical depth can be calculated analytically as
\begin{equation}
  \tau_{\gamma\gamma} = \frac{4L_{\mathrm{ICL}} \sigma_{\mathrm{T}}} {3R_{\mathrm{ICL}} c 
  \epsilon_{\mathrm{eff}}^2 m_{\mathrm{e}}c^2} \frac{2}{\epsilon_{\gamma}} 
  \delta(\epsilon_{\gamma} - \frac{2}{\epsilon_{\mathrm{eff}}}).
\end{equation}
For the characteristic \gammaray\ energy 
\begin{equation}
  E_{\mathrm{\gamma}} = \frac{2}{\epsilon} m_{\mathrm{e}} c^2 = 522 E_{\mathrm{eV}}^{-1}\,\mathrm{GeV},
  \label{Egamma}
\end{equation} 
where $E_{\mathrm{eV}}$ is the energy of the target photon in eV, the \gaga\ optical 
depth can then be estimated as 
\begin{align}
\tau_{\gamma\gamma}(E_{\mathrm{eV}}) &= 2.4 \times 10^{-12} E_{\mathrm{eV}}^{-1} \left(
\frac{L_{\mathrm{ICL}}}{L_{\odot}}\right) \left(\frac{R_{\mathrm{ICL}}}{10\,\mathrm{kpc}}\right)^{-1} \\
&= 2.4 \times 10^{\frac{M_\odot - M_{\mathrm{ICL}}}{2.5} - 12} 
E_{\mathrm{eV}}^{-1}  \left(\frac{R_{\mathrm{ICL}}}{10\,\mathrm{kpc}}\right)^{-1}. 
\end{align}

In a search for ICL in a sample of ten clusters at redshifts $0.4 < z < 0.8$, \citet{2012A&A...537A..64G}
detected diffuse light in all ten clusters with typical sizes of a few tens of kpc and a total 
ICL magnitude in the range from -18 to -21. Those authors also show that there are no strong 
variations in the amount of ICL between $z=0$ and $z = 0.8$ with just a modest increase 
\citep{2012A&A...537A..64G}. HST ACS\footnote{The Advanced Camera for Surveys (ACS) aboard the Hubble Space Telescope (HST)} 
images in the F814W filter were used in this study. 
The HST F814W filter covers a wavelength range from $6948$ \AA\ to 10043 \AA\ with an 
effective wavelength of $\lambda_{\mathrm{eff}} = 8186.4$\,\AA\ which corresponds to an energy 
of $E_\mathrm{eff} = 1.5$\,eV. Using the lowest measured value of the absolute magnitude of the
ICL and a size of the ICL region of 10 kpc, we can estimate an upper limit on the ICL optical 
depth as
\begin{equation}
\label{opacity}
\tau_{\gamma\gamma} < 4.8 \times 10^{-2} E_{\mathrm{eV}}^{-1}.
\end{equation}
However, for a more precise estimate of the ICL luminosity, the absolute ICL magnitude 
should be corrected by the bolometric correction which is dependent on the effective 
temperature of the source \citep{2008PASP..120..583G}. Unfortunately, the analysis 
performed in \citet{2012A&A...537A..64G} was unable to determine colours of the 
detected ICL sources, but e.g. \citet{2005A&A...429...39A} determine the colours 
of the diffuse light sources in the Coma cluster to correspond to quite old stellar 
populations. For an effective temperature in the range of $\sim4000-6000$\,K the 
bolometric correction for the F814W filter is estimated to be roughly $0.5$ 
\citep{2008PASP..120..583G}, which would decrease the upper limit for the optical 
depth even further. Therefore, we can firmly conclude that even in the most favourable 
configuration, the \gaga\,absorption due to the ICL is negligible.

\section{\label{galaxies}Nearby galaxies}

The \gaga\,absorption of the blazar emission due to an intervening galaxy close to the 
line of sight between the observer and a blazar is discussed in detail in \citet{2014ApJ...790..147B}. 
They show that only in the case of a very compact and, at the same time, luminous galaxy 
and a small value of the impact parameter (distance of closest approach of the $\gamma$-ray
trajectory to the nucleus of the galaxy) of $\lesssim 10 R_{\mathrm{eff}}$, where 
$R_{\mathrm{eff}}$ is the effective radius of the galaxy, is substantial \gaga\,absorption 
possible. However, they also note that galaxies of such high luminosities are typically 
giant ellipticals or large spirals with larger effective radii than 1 kpc and, thus, the 
expected \gaga\ absorption from the radiation field of an intervening galaxy is always
neglible \citep{2014ApJ...790..147B}. 

However, in cluster environments it is possible that a blazar is surrounded by several companion 
galaxies. In this case, the combined radiation field of all nearby galaxies should be considered, 
and it might still provide a substantial \gaga-absorption opacity. The optical depth of the combined 
radiation field can be estimated as 
\begin{equation}
\tau_{\gamma\gamma} = \int dx \int d\Omega \, (1-\mu) \int d\epsilon \, n_\mathrm{tot}(\epsilon,\Omega,x)
\, \sigma(\epsilon, \epsilon_{\gamma}, \mu)
\end{equation}
where $n_\mathrm{tot}(\epsilon,\Omega,x)$ is the total differential photon density of the 
combined radiation field of all the nearby galaxies. The total differential photon density 
is the superposition of the photon densities of the radiation fields created by individual 
galaxies, i.e. 
\begin{equation}
  n_\mathrm{tot}(\epsilon,\Omega,x) = \sum_{i = 1}^{N} n_i(\epsilon,\Omega,x),
\end{equation} 
where $N$ is the number of nearby galaxies. Therefore the optical depth of the combined 
radiation field can also be represented as the sum of the optical depths due to the individual
radiation fields from individual galaxies, i.e. 
\begin{equation}
  \tau_{\gamma\gamma} = \sum_{i = 1}^{N} \tau_{\gamma\gamma}^i.
  \label{tau_sum}
\end{equation} 

In order to calculate the optical depth due to the radiation field of an individual galaxy 
and to study its dependence on the location of the galaxy with respect to the blazar, its 
luminosity and effective temperature, we followed the assumptions presented in the Appendix 
of \citet{2014ApJ...790..147B}: the galaxy is approximated by a flat disk with a De Vaucouleurs 
surface brightness profile, and the spectrum of the galaxy is approximated by black-body 
radiation. The inclination of the disk is characterazed by the inclination angle $i$ between 
the disk normal vector and the line of sight. \citet{2014ApJ...790..147B} showed that the 
optical depth is only weakly dependent on the value of the inclination angle, so in the 
calculations performed in this paper we assume $i = 0^\circ$. The integration is perfomed 
numerically, properly accounting for the angular dependence.

\begin{figure}
\centering
\resizebox{\hsize}{!}{\includegraphics{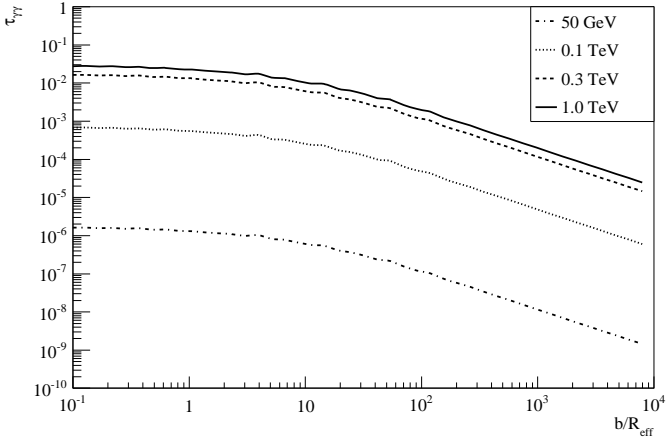}}
\caption{Gamma-gamma optical depth of a radiation field of a companion galaxy 
as a function of the impact parameter $b$ presented for different \gammaray\ energies, 
assuming the galaxy is located at the 
same distance. The parameters of the galaxy are assumed to be $R_\mathrm{eff} = 1$\,kpc, 
$L = L_\mathrm{MW}$, and $T = 5000$\,K.}
\label{tau_b}
\end{figure}

\begin{figure}
\centering
\resizebox{\hsize}{!}{\includegraphics{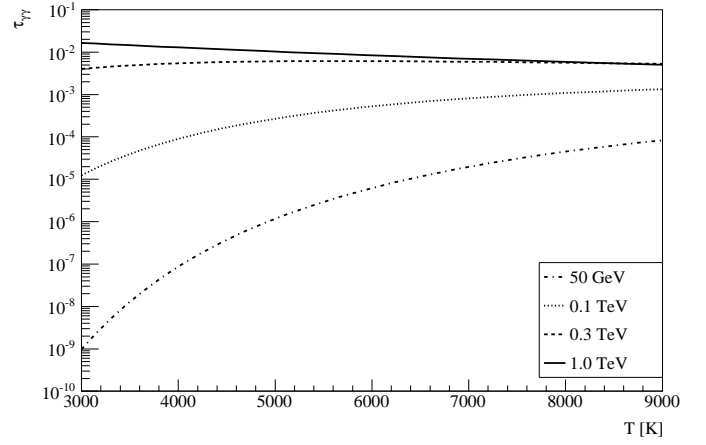}}
\caption{The same as Figure\,\ref{tau_b}, but as a function of temperature $T$ for the impact parameter of $10R_\mathrm{eff}$.}
\label{tau_T}
\end{figure}

Figure \ref{tau_b} shows the resulting \gaga\ optical depth for $\gamma$\,rays passing through 
the radiation field of a companion galaxy as a function of the impact parameter $b$ for 
different \gammaray\ energies. The galaxy is assumed to have the same same luminosity as 
the Milky Way, $L_\mathrm{MW} = 9.2\times10^{43}$\,erg/s, with an effective radius of 
$R_\mathrm{eff} = 1$\,kpc and effective temperature of $T=5000$\,K. The figure shows that 
for such an individual companion galaxy the \gaga\,absorption is negligible. To cause
significant absorption a luminosity of $L\gtrsim 10L_\mathrm{MW}$ is needed for the 
galaxy with the same parameters because the optical depth is linearly dependent on luminosity. 
The dependence of the \gaga\ optical depth on the effective temperature of the galaxy is 
shown in Figure\,\ref{tau_T}. According to the resonance condition in Eq. (\ref{Egamma}),
with $E_{\rm eff} \approx 2.8 \, k T$, photons of $E_{\gamma} = 1.1 \, T_{5000}^{-1}$~TeV
are most efficiently absorbed by photons emitted by a black-body emitter at $T = 5000 \,
T_{5000}$~K. Hence, photons with energies below $\sim 1$~TeV are approaching this condition
with increasing black-body temperature, so the optical depth increases. 

\begin{figure}
\centering
\resizebox{\hsize}{!}{\includegraphics{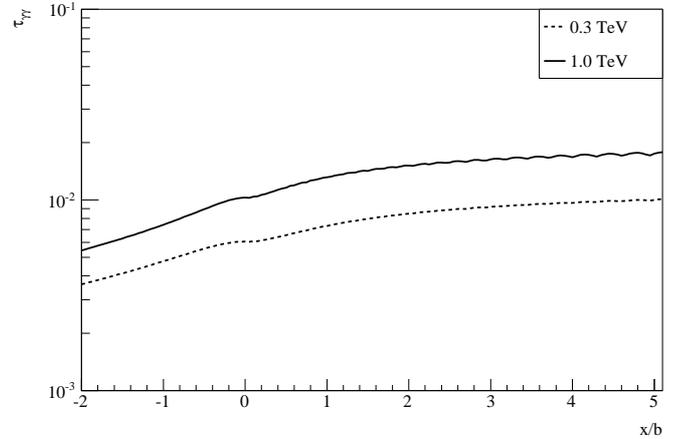}}
\caption{The same as Figure\,\ref{tau_b}, but as a function of the location of the companion galaxy with respect to the source of the \gammaray\ emission for the impact parameter of $10R_\mathrm{eff}$.}
\label{tau_x}
\end{figure}

Equation\,\ref{tau_sum} suggests that \gaga\,absorption might become substantial when the 
blazar is surrounded by $\gtrsim 5$ companion galaxies. The optical depth also depends on 
the location of the companion galaxy with respect to the observer. Obviously, the absorption 
will be more effective when the companion galaxy is located in front (as seen from the observer
on Earth) rather than behind the blazar. To calculate the dependence of the optical depth on 
the location of the galaxy with the respect to the observer we define the x-axis with an 
origin in the blazar and aligned with the jet, assuming that the jet is pointing towards 
the observer. The resulting dependence of the optical depth on $x$ is illustrated in 
Fig.\,\ref{tau_x}

\section{\label{1440}\source}

For a search of possible candidates of blazars with nearby companions, we used the HST survey of BL Lacertae objects 
\citep{2000ApJ...532..740S, 2000ApJ...532..816U, 2000ApJ...542..731F, 2000ApJ...544..258S} in which 
110 objects were observed. For a subset of 30 nearby ($z\lesssim0.2$) BL Lac objects with the highest 
signal-to-noise ratios a morphological sudy was performed, including the investigation of the near 
environment and a search for companions \citep{2000ApJ...542..731F}. For 11 out of the 30 objects close 
companions were found. For three of them (0706+592, 1440+122, and 2356$-$309) very close 
($\delta<1.2^{\prime\prime}$) compact companions were detected located at projected distances 
of 1 -- 5~kpc from the nucleus, assuming the same redshift as the BL Lac object. For four objects 
(0829+046, 1229+645, 1440+122, and 1853+671) companion galaxies were detected located at projected 
distances of 10 --20~kpc from the BL Lac object. Among these objects, there are only three which 
are detected at both GeV \citep{2011ApJ...743..171A} and TeV energies: 0706+592 
\citep{2010ApJ...715L..49A}, 1440+122 \citep{2011arXiv1110.0040W}, and 2356$-$309 
\citep{2006A&A...455..461A}. Only one of these has a companion galaxy: 1440+122. 
Two others have relatively faint compact companions with absolute magnitudes of 
$-15$ and $-17.3$. Therefore, we have chosen 1ES\,1440+122 for a case study, as a primary 
candidate for possible \gaga\,absorption because of the nearby companions.

\begin{figure}
\centering
\resizebox{\hsize}{!}{\includegraphics{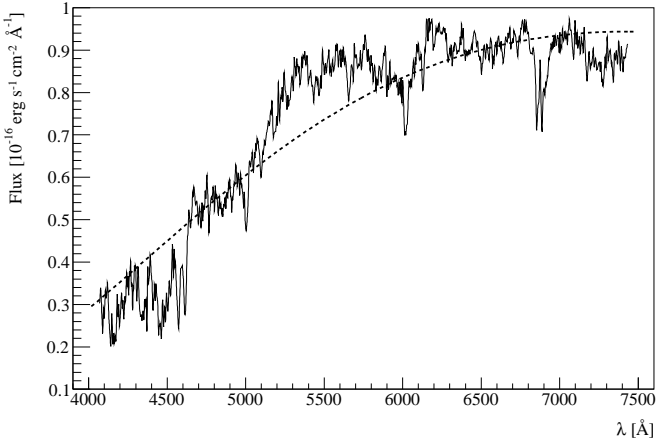}}
\caption{A spectrum of a companion galaxy of \source\ (solid line) \citep{2006A&A...457...35S} approximated by a Planckian distribution (dashed line).}
\label{spectrum}
\end{figure}

The source \source\ is a BL Lac type object detected in all energy bands across the electro-magnetic 
spectrum, from radio to VHE $\gamma$\,rays. It is located at a redshift $z = 0.162$ 
\citep{1993ApJ...412..541S} and appears to be located in a rich environment, being 
surrounded by $\sim 20$ galaxies \citep{1999A&A...341..683H}. According to the HST 
survey \citep{2000ApJ...542..731F}, \source\ features two close companions: a galaxy at an 
angular distance of $2.5^{\prime\prime}$ and a very close compact companion at the distance 
of $0.3^{\prime\prime}$. \citet{2004ApJ...613..747G} showed that the close compact companion 
of \source\ is a foreground star and therefore it is not of interest for this paper. An optical 
spectroscopy study \citep{2006A&A...457...35S} revealed a spectrum of the close companion galaxy 
classifying it as an elliptical galaxy at a redshift of $z = 0.161$, showing that 
these objects are located in the same cluster. The projected distance between \source\ and 
the companion galaxy is $\simeq13$\,kpc. The R-band apparent magnitude of the companion galaxy 
is $m_{\mathrm{R}} = 17.2$ \citep{2006A&A...457...35S} which corresponds to an absolute magnitude 
in the R-band of $M_{\mathrm{R}} = -23.4$. This corresponds, neglecting the bolometric correction, 
to a luminosity of $\simeq 8\times10^{44}$\,erg/s, which is about 10 times higher than the 
luminosity of the Milky Way. The effective temperature of the galaxy is estimated to be 
$T \simeq 3900$\,K using a fit of a Planckian distribution to the spectrum of the galaxy obtained in 
\citet{2006A&A...457...35S} (Figure \ref{spectrum}).

\begin{figure}
\centering
\resizebox{\hsize}{!}{\includegraphics{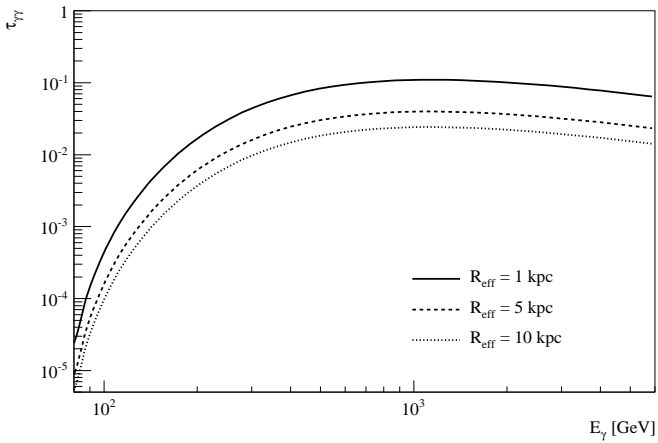}}
\caption{A \gaga\ opacity of the $\gamma$\,ray emitted from \source\ in the radiation field of the 
companion galaxy as a function of the \gammaray\ energy for various values of the effective radius 
of the companion galaxy.}
\label{tau_E_1440}
\end{figure}

In Figure \ref{tau_E_1440}, the optical depth of the \gaga\,absorption caused by the radiation 
field of the companion galaxy of \source\ is shown as a function of \gammaray\ energy for 
different values of the effective radius of the galaxy. Substantial absorption is possible 
only in the case of a relatively small effective radius of the galaxy, $R_\mathrm{eff}\lesssim1$\,kpc. 
However, the typical size of such luminous galaxies is typically much bigger, of the order
of $R_{\mathrm{eff}}\gtrsim10$\,kpc. Therefore, significant absorption by the radiation field
of this individual galaxy is very unlikely.

\section{\label{cascades}Cascade emission}

As mentioned above, if a VHE $\gamma$-ray blazar is surrounded by $\gtrsim 5$ luminous 
galaxies \gaga\,absorption might become substantial. Although we failed to find examples 
of such dense environments with multiple companion galaxies, such environments might exist 
at higher redshifts of $z \gtrsim 2$ where cluster environments are believed to have been 
denser than at the present epoch. However, the VHE \gammaray\ emission from blazars at 
$z \gtrsim 2$ is heavily absorbed by the EBL and cannot be detected. Therefore, it is impossible 
to examine blazar environments directly by studying TeV emission. Nevertheless, absorbed 
$\gamma$\,rays might be re-emitted at GeV energies through Compton-supported, \gammaray-induced 
pair cascades and contribute to the observed GeV emission. Below we provide simple estimates
of the flux level of the expected GeV emission from \gammaray-induced pair cascades in the 
combined photon field of several companion galaxies.

Based on the example of \source\ which is neighbours a luminous elliptical galaxy, the maximum 
optical depth in the radiation field of a single nearby galaxy is $\tau_{\mathrm{max}} \simeq 0.02$. 
For $N$ nearby galaxies with similar properties the maximum optical depth is then 
$\tau_\mathrm{max}^{\mathrm{tot}}\simeq N \tau_{\mathrm{max}}$. Assuming that all energy in
the absorbed \gammaray\ emission is re-emitted at GeV energies, we can find the most optimistic
estimate for the flux from the cascade emission in the limit that the magnetic field is weak 
enough not to deflect the electrons/positrons in the cascade from the primary direction of the
$\gamma$\,rays. In that case, the secondary $\gamma$\,rays are re-emitted in the same direction and
with the same beaming characteristic as the original VHE $\gamma$-ray beam, and the expected 
cascade emission in the GeV energy band is 
\begin{equation}
  F_{\mathrm{casc}}^{\mathrm{GeV}} \simeq (1 - e^{-\tau_\mathrm{max}^{\mathrm{tot}}})
  F_{\mathrm{int}}^{\mathrm{TeV}} \simeq \tau_\mathrm{max}^{\mathrm{tot}} 
  F_{\mathrm{int}}^{\mathrm{TeV}},
\end{equation} 
where the intrinsic TeV flux from the blazar jet can be approximated by 
\begin{equation}
F_{\mathrm{int}}^{\mathrm{TeV}} \simeq \frac{1}{2}\frac{L}{4\pi d_{\mathrm{L}}^2 },
\end{equation}
where $d_{\mathrm{L}}$ is the luminosity distance to the blazar and the factor of $1/2$ accounts for the
presence of two jets; $L = 10^{44} \, L_{44}$~erg~s$^{-1}$ is the inferred isotropic VHE $\gamma$-ray 
luminosity which usually does not exceed $\sim 10^{44}$~erg~s$^{-1}$ for known VHE $\gamma$-ray blazars.
In the case of the substantial magnetic field ($B \gtrsim 10^{-14}$~G), electrons/positrons in the cascade
may be deflected out of the primary VHE $\gamma$-ray beam, and the GeV $\gamma$\,rays will be re-distributed 
over a larger solid angle, resulting in a lower flux. 

For two different assumed values of the redshift $z = 0.1$ and $z = 2.0$, the upper limits 
for the observed cascade GeV emission are
\begin{align}
  z = 0.1: & \,\,\,F_{\mathrm{casc}}^{\mathrm{GeV}} \simeq 2.1 \times 10^{-14} L_{44} N \,\mathrm{erg\,cm}^{-2}\mathrm{s}^{-1},\\
  z = 2.0: & \,\,\,F_{\mathrm{casc}}^{\mathrm{GeV}} \simeq 1.8 \times 10^{-17} L_{44} N \,\mathrm{erg\,cm}^{-2}\mathrm{s}^{-1}.
\end{align} 
Comparing this to the \fermi\ sensitivity of $F_{\rm min}^{\rm LAT} \sim 10^{-12}$~erg~cm$^{-2}$~s$^{-1}$,
it is obvious that even for $\sim 10$~nearby, luminous companion galaxies and a large intrinsic
VHE luminosity of the high-redshift blazar, the expected cascade flux is not expected to make a
measurable contribution to the GeV $\gamma$-ray flux from galaxy clusters hosting radio-loud AGN.

\section{\label{summary}Summary}

We have evaluated the optical depth for VHE $\gamma$\,rays produced in blazars due to $\gamma\gamma$ 
absorption in the cluster environments of the blazar's host galaxy. Considering target photon 
fields from the intracluster light and companion galaxies within the cluster, we conclude that
neither of these fields is likely to constitute a substantial opacity to $\gamma\gamma$ absorption,
unless a large number ($\gtrsim 5$) of very luminous, compact galaxies are located close to the
line of sight of the blazar beam. This situation is not found in the local Universe, but may
be possible at larger redshifts ($z \gtrsim 2$). Since VHE $\gamma$\,rays from such distances are 
expected to be heavily attenuated by $\gamma\gamma$ absorption by the EBL, we considered the
possible signatures of VHE $\gamma$-ray induced, Compton-supported pair cascades following the
$\gamma\gamma$ absorption of blazar emission in dense cluster environments, but concluded that
the expected GeV flux level is unlikely to be detectable by \fermi.

\acknowledgements{The authors thank C. Burke for stimulating discussions on the Intracluster 
Light. MB acknowledges support from the South African Department of Science and Technology 
through the National Research Foundation under NRF SARChI Chair grant No. 64789.}

\bibliographystyle{aa} 
\bibliography{cluster_absorption_aanda_arxiv.bbl}

\begin{thebibliography}{25}
\expandafter\ifx\csname natexlab\endcsname\relax\def\natexlab#1{#1}\fi

\bibitem[{{Acciari} {et~al.}(2010){Acciari}, {Aliu}, {Arlen}, {Aune},
  {Bautista}, {Beilicke}, {Benbow}, {B{\"o}ttcher}, {Boltuch}, {Bradbury}, \&
  et~al.}]{2010ApJ...715L..49A}
{Acciari}, V.~A., {Aliu}, E., {Arlen}, T., {et~al.} 2010, \apjl, 715, L49

\bibitem[{{Ackermann} {et~al.}(2011){Ackermann}, {Ajello}, {Allafort},
  {Antolini}, {Atwood}, {Axelsson}, {Baldini}, {Ballet}, {Barbiellini},
  {Bastieri}, {Bechtol}, {Bellazzini}, {Berenji}, {Blandford}, {Bloom},
  {Bonamente}, {Borgland}, {Bottacini}, {Bouvier}, {Bregeon}, {Brigida},
  {Bruel}, {Buehler}, {Burnett}, {Buson}, {Caliandro}, {Cameron}, {Caraveo},
  {Casandjian}, {Cavazzuti}, {Cecchi}, {Charles}, {Cheung}, {Chiang},
  {Ciprini}, {Claus}, {Cohen-Tanugi}, {Conrad}, {Costamante}, {Cutini}, {de
  Angelis}, {de Palma}, {Dermer}, {Digel}, {Silva}, {Drell}, {Dubois},
  {Escande}, {Favuzzi}, {Fegan}, {Ferrara}, {Finke}, {Focke}, {Fortin},
  {Frailis}, {Fukazawa}, {Funk}, {Fusco}, {Gargano}, {Gasparrini}, {Gehrels},
  {Germani}, {Giebels}, {Giglietto}, {Giommi}, {Giordano}, {Giroletti},
  {Glanzman}, {Godfrey}, {Grenier}, {Grove}, {Guiriec}, {Gustafsson},
  {Hadasch}, {Hayashida}, {Hays}, {Healey}, {Horan}, {Hou}, {Hughes},
  {Iafrate}, {J{\'o}hannesson}, {Johnson}, {Johnson}, {Kamae}, {Katagiri},
  {Kataoka}, {Kn{\"o}dlseder}, {Kuss}, {Lande}, {Larsson}, {Latronico},
  {Longo}, {Loparco}, {Lott}, {Lovellette}, {Lubrano}, {Madejski}, {Mazziotta},
  {McConville}, {McEnery}, {Michelson}, {Mitthumsiri}, {Mizuno}, {Moiseev},
  {Monte}, {Monzani}, {Moretti}, {Morselli}, {Moskalenko}, {Murgia},
  {Nakamori}, {Naumann-Godo}, {Nolan}, {Norris}, {Nuss}, {Ohno}, {Ohsugi},
  {Okumura}, {Omodei}, {Orienti}, {Orlando}, {Ormes}, {Ozaki}, {Paneque},
  {Parent}, {Pesce-Rollins}, {Pierbattista}, {Piranomonte}, {Piron}, {Pivato},
  {Porter}, {Rain{\`o}}, {Rando}, {Razzano}, {Razzaque}, {Reimer}, {Reimer},
  {Ritz}, {Rochester}, {Romani}, {Roth}, {Sanchez}, {Sbarra}, {Scargle},
  {Schalk}, {Sgr{\`o}}, {Shaw}, {Siskind}, {Spandre}, {Spinelli}, {Strong},
  {Suson}, {Tajima}, {Takahashi}, {Takahashi}, {Tanaka}, {Thayer}, {Thayer},
  {Thompson}, {Tibaldo}, {Tinivella}, {Torres}, {Tosti}, {Troja}, {Uchiyama},
  {Vandenbroucke}, {Vasileiou}, {Vianello}, {Vitale}, {Waite}, {Wallace},
  {Wang}, {Winer}, {Wood}, {Wood}, \& {Zimmer}}]{2011ApJ...743..171A}
{Ackermann}, M., {Ajello}, M., {Allafort}, A., {et~al.} 2011, \apj, 743, 171

\bibitem[{{Adami} {et~al.}(2005){Adami}, {Slezak}, {Durret}, {Conselice},
  {Cuillandre}, {Gallagher}, {Mazure}, {Pell{\'o}}, {Picat}, \&
  {Ulmer}}]{2005A&A...429...39A}
{Adami}, C., {Slezak}, E., {Durret}, F., {et~al.} 2005, \aap, 429, 39

\bibitem[{{Aharonian} {et~al.}(2006){Aharonian}, {Akhperjanian}, {Bazer-Bachi},
  {Beilicke}, {Benbow}, {Berge}, {Bernl{\"o}hr}, {Boisson}, {Bolz}, {Borrel},
  {Braun}, {Breitling}, {Brown}, {B{\"u}hler}, {B{\"u}sching}, {Carrigan},
  {Chadwick}, {Chounet}, {Cornils}, {Costamante}, {Degrange}, {Dickinson},
  {Djannati-Ata{\"i}}, {O'C.~Drury}, {Dubus}, {Egberts}, {Emmanoulopoulos},
  {Espigat}, {Feinstein}, {Ferrero}, {Fontaine}, {Funk}, {Funk}, {Gallant},
  {Giebels}, {Glicenstein}, {Goret}, {Hadjichristidis}, {Hauser}, {Hauser},
  {Heinzelmann}, {Henri}, {Hermann}, {Hinton}, {Hofmann}, {Holleran}, {Horns},
  {Jacholkowska}, {de Jager}, {Kh{\'e}lifi}, {Komin}, {Konopelko}, {Latham},
  {Le Gallou}, {Lemi{\`e}re}, {Lemoine-Goumard}, {Lohse}, {Martin},
  {Martineau-Huynh}, {Marcowith}, {Masterson}, {McComb}, {de Naurois},
  {Nedbal}, {Nolan}, {Noutsos}, {Orford}, {Osborne}, {Ouchrif}, {Panter},
  {Pelletier}, {Pita}, {P{\"u}hlhofer}, {Punch}, {Raubenheimer}, {Raue},
  {Rayner}, {Reimer}, {Reimer}, {Ripken}, {Rob}, {Rolland}, {Rowell},
  {Sahakian}, {Saug{\'e}}, {Schlenker}, {Schlickeiser}, {Schwanke}, {Sol},
  {Spangler}, {Spanier}, {Steenkamp}, {Stegmann}, {Superina}, {Tavernet},
  {Terrier}, {Th{\'e}oret}, {Tluczykont}, {van Eldik}, {Vasileiadis}, {Venter},
  {Vincent}, {V{\"o}lk}, {Wagner}, \& {Ward}}]{2006A&A...455..461A}
{Aharonian}, F., {Akhperjanian}, A.~G., {Bazer-Bachi}, A.~R., {et~al.} 2006,
  \aap, 455, 461

\bibitem[{{Barnacka} {et~al.}(2014){Barnacka}, {B{\"o}ttcher}, \&
  {Sushch}}]{2014ApJ...790..147B}
{Barnacka}, A., {B{\"o}ttcher}, M., \& {Sushch}, I. 2014, \apj, 790, 147

\bibitem[{{Benbow} {et~al.}(2011)}]{2011arXiv1110.0040W}
{Benbow}, W. {for the VERITAS Collaboration}. 2011, in Proceedings of the
  32nd International Cosmic Ray Conference, [arXiv:1110.0040]

\bibitem[{{Burke} {et~al.}(2012){Burke}, {Collins}, {Stott}, \&
  {Hilton}}]{2012MNRAS.425.2058B}
{Burke}, C., {Collins}, C.~A., {Stott}, J.~P., \& {Hilton}, M. 2012, \mnras,
  425, 2058

\bibitem[{{Falomo} {et~al.}(2000){Falomo}, {Scarpa}, {Treves}, \&
  {Urry}}]{2000ApJ...542..731F}
{Falomo}, R., {Scarpa}, R., {Treves}, A., \& {Urry}, C.~M. 2000, \apj, 542, 731

\bibitem[{{Finke} {et~al.}(2010){Finke}, {Razzaque}, \&
  {Dermer}}]{2010ApJ...712..238F}
{Finke}, J.~D., {Razzaque}, S., \& {Dermer}, C.~D. 2010, \apj, 712, 238

\bibitem[{{Franceschini} {et~al.}(2008){Franceschini}, {Rodighiero}, \&
  {Vaccari}}]{2008A&A...487..837F}
{Franceschini}, A., {Rodighiero}, G., \& {Vaccari}, M. 2008, \aap, 487, 837

\bibitem[{{Giovannini} {et~al.}(2004){Giovannini}, {Falomo}, {Scarpa},
  {Treves}, \& {Urry}}]{2004ApJ...613..747G}
{Giovannini}, G., {Falomo}, R., {Scarpa}, R., {Treves}, A., \& {Urry}, C.~M.
  2004, \apj, 613, 747

\bibitem[{{Girardi} {et~al.}(2008){Girardi}, {Dalcanton}, {Williams}, {de
  Jong}, {Gallart}, {Monelli}, {Groenewegen}, {Holtzman}, {Olsen}, {Seth},
  {Weisz}, \& {ANGST/ANGRRR Collaboration}}]{2008PASP..120..583G}
{Girardi}, L., {Dalcanton}, J., {Williams}, B., {et~al.} 2008, \pasp, 120, 583

\bibitem[{{Gould} \& {Schr{\'e}der}(1967)}]{1967PhRv..155.1404G}
{Gould}, R.~J. \& {Schr{\'e}der}, G.~P. 1967, Physical Review, 155, 1404

\bibitem[{{Guennou} {et~al.}(2012){Guennou}, {Adami}, {Da Rocha}, {Durret},
  {Ulmer}, {Allam}, {Basa}, {Benoist}, {Biviano}, {Clowe}, {Gavazzi},
  {Halliday}, {Ilbert}, {Johnston}, {Just}, {Kron}, {Kubo}, {Le Brun},
  {Marshall}, {Mazure}, {Murphy}, {Pereira}, {Raba{\c c}a}, {Rostagni},
  {Rudnick}, {Russeil}, {Schrabback}, {Slezak}, {Tucker}, \&
  {Zaritsky}}]{2012A&A...537A..64G}
{Guennou}, L., {Adami}, C., {Da Rocha}, C., {et~al.} 2012, \aap, 537, A64

\bibitem[{{Heidt} {et~al.}(1999){Heidt}, {Nilsson}, {Sillanp{\"a}{\"a}},
  {Takalo}, \& {Pursimo}}]{1999A&A...341..683H}
{Heidt}, J., {Nilsson}, K., {Sillanp{\"a}{\"a}}, A., {Takalo}, L.~O., \&
  {Pursimo}, T. 1999, \aap, 341, 683

\bibitem[{{Planck Collaboration} {et~al.}(2014){Planck Collaboration}, {Ade, P.
  A. R.}, {Aghanim, N.}, {Armitage-Caplan, C.}, {Arnaud, M.}, {Ashdown, M.},
  {Atrio-Barandela, F.}, {Aumont, J.}, {Baccigalupi, C.}, {Banday, A. J.},
  {Barreiro, R. B.}, {Bartlett, J. G.}, {Battaner, E.}, {Benabed, K.},
  {BenoÃ®t, A.}, {Benoit-LÃ©vy, A.}, {Bernard, J.-P.}, {Bersanelli, M.},
  {Bielewicz, P.}, {Bobin, J.}, {Bock, J. J.}, {Bonaldi, A.}, {Bond, J. R.},
  {Borrill, J.}, {Bouchet, F. R.}, {Bridges, M.}, {Bucher, M.}, {Burigana, C.},
  {Butler, R. C.}, {Calabrese, E.}, {Cappellini, B.}, {Cardoso, J.-F.},
  {Catalano, A.}, {Challinor, A.}, {Chamballu, A.}, {Chary, R.-R.}, {Chen, X.},
  {Chiang, H. C.}, {Chiang, L.-Y}, {Christensen, P. R.}, {Church, S.},
  {Clements, D. L.}, {Colombi, S.}, {Colombo, L. P. L.}, {Couchot, F.},
  {Coulais, A.}, {Crill, B. P.}, {Curto, A.}, {Cuttaia, F.}, {Danese, L.},
  {Davies, R. D.}, {Davis, R. J.}, {de Bernardis, P.}, {de Rosa, A.}, {de
  Zotti, G.}, {Delabrouille, J.}, {Delouis, J.-M.}, {DÃ©sert, F.-X.},
  {Dickinson, C.}, {Diego, J. M.}, {Dolag, K.}, {Dole, H.}, {Donzelli, S.},
  {DorÃ©, O.}, {Douspis, M.}, {Dunkley, J.}, {Dupac, X.}, {Efstathiou, G.},
  {Elsner, F.}, {EnÃlin, T. A.}, {Eriksen, H. K.}, {Finelli, F.}, {Forni,
  O.}, {Frailis, M.}, {Fraisse, A. A.}, {Franceschi, E.}, {Gaier, T. C.},
  {Galeotta, S.}, {Galli, S.}, {Ganga, K.}, {Giard, M.}, {Giardino, G.},
  {Giraud-HÃ©raud, Y.}, {GjerlÃ¸w, E.}, {GonzÃ¡lez-Nuevo, J.},
  {GÃ³rski, K. M.}, {Gratton, S.}, {Gregorio, A.}, {Gruppuso, A.},
  {Gudmundsson, J. E.}, {Haissinski, J.}, {Hamann, J.}, {Hansen, F. K.},
  {Hanson, D.}, {Harrison, D.}, {Henrot-VersillÃ©, S.},
  {HernÃ¡ndez-Monteagudo, C.}, {Herranz, D.}, {Hildebrandt, S. R.}, {Hivon,
  E.}, {Hobson, M.}, {Holmes, W. A.}, {Hornstrup, A.}, {Hou, Z.}, {Hovest, W.},
  {Huffenberger, K. M.}, {Jaffe, A. H.}, {Jaffe, T. R.}, {Jewell, J.}, {Jones,
  W. C.}, {Juvela, M.}, {KeihÃ¤nen, E.}, {Keskitalo, R.}, {Kisner, T. S.},
  {Kneissl, R.}, {Knoche, J.}, {Knox, L.}, {Kunz, M.}, {Kurki-Suonio, H.},
  {Lagache, G.}, {LÃ¤hteenmÃ¤ki, A.}, {Lamarre, J.-M.}, {Lasenby, A.},
  {Lattanzi, M.}, {Laureijs, R. J.}, {Lawrence, C. R.}, {Leach, S.}, {Leahy, J.
  P.}, {Leonardi, R.}, {LeÃ³n-Tavares, J.}, {Lesgourgues, J.}, {Lewis, A.},
  {Liguori, M.}, {Lilje, P. B.}, {Linden-VÃ¸rnle, M.}, {LÃ³pez-Caniego,
  M.}, {Lubin, P. M.}, {MacÃ­as-PÃ©rez, J. F.}, {Maffei, B.}, {Maino, D.},
  {Mandolesi, N.}, {Maris, M.}, {Marshall, D. J.}, {Martin, P. G.},
  {MartÃ­nez-GonzÃ¡lez, E.}, {Masi, S.}, {Massardi, M.}, {Matarrese, S.},
  {Matthai, F.}, {Mazzotta, P.}, {Meinhold, P. R.}, {Melchiorri, A.}, {Melin,
  J.-B.}, {Mendes, L.}, {Menegoni, E.}, {Mennella, A.}, {Migliaccio, M.},
  {Millea, M.}, {Mitra, S.}, {Miville-DeschÃªnes, M.-A.}, {Moneti, A.},
  {Montier, L.}, {Morgante, G.}, {Mortlock, D.}, {Moss, A.}, {Munshi, D.},
  {Murphy, J. A.}, {Naselsky, P.}, {Nati, F.}, {Natoli, P.}, {Netterfield, C.
  B.}, {NÃ¸rgaard-Nielsen, H. U.}, {Noviello, F.}, {Novikov, D.}, {Novikov,
  I.}, {OâDwyer, I. J.}, {Osborne, S.}, {Oxborrow, C. A.}, {Paci, F.},
  {Pagano, L.}, {Pajot, F.}, {Paladini, R.}, {Paoletti, D.}, {Partridge, B.},
  {Pasian, F.}, {Patanchon, G.}, {Pearson, D.}, {Pearson, T. J.}, {Peiris, H.
  V.}, {Perdereau, O.}, {Perotto, L.}, {Perrotta, F.}, {Pettorino, V.},
  {Piacentini, F.}, {Piat, M.}, {Pierpaoli, E.}, {Pietrobon, D.},
  {Plaszczynski, S.}, {Platania, P.}, {Pointecouteau, E.}, {Polenta, G.},
  {Ponthieu, N.}, {Popa, L.}, {Poutanen, T.}, {Pratt, G. W.}, {PrÃ©zeau, G.},
  {Prunet, S.}, {Puget, J.-L.}, {Rachen, J. P.}, {Reach, W. T.}, {Rebolo, R.},
  {Reinecke, M.}, {Remazeilles, M.}, {Renault, C.}, {Ricciardi, S.}, {Riller,
  T.}, {Ristorcelli, I.}, {Rocha, G.}, {Rosset, C.}, {Roudier, G.},
  {Rowan-Robinson, M.}, {RubiÃ±o-MartÃ­n, J. A.}, {Rusholme, B.}, {Sandri,
  M.}, {Santos, D.}, {Savelainen, M.}, {Savini, G.}, {Scott, D.}, {Seiffert, M.
  D.}, {Shellard, E. P. S.}, {Spencer, L. D.}, {Starck, J.-L.}, {Stolyarov,
  V.}, {Stompor, R.}, {Sudiwala, R.}, {Sunyaev, R.}, {Sureau, F.}, {Sutton,
  D.}, {Suur-Uski, A.-S.}, {Sygnet, J.-F.}, {Tauber, J. A.}, {Tavagnacco, D.},
  {Terenzi, L.}, {Toffolatti, L.}, {Tomasi, M.}, {Tristram, M.}, {Tucci, M.},
  {Tuovinen, J.}, {TÃ¼rler, M.}, {Umana, G.}, {Valenziano, L.}, {Valiviita,
  J.}, {Van Tent, B.}, {Vielva, P.}, {Villa, F.}, {Vittorio, N.}, {Wade, L.
  A.}, {Wandelt, B. D.}, {Wehus, I. K.}, {White, M.}, {White, S. D. M.},
  {Wilkinson, A.}, { Yvon, D.}, {Zacchei, A.}, \& {Zonca, A.}}]{planck}
{Planck Collaboration}, {Ade, P. A. R.}, {Aghanim, N.}, {et~al.} 2014, \aap,
  571, A16

\bibitem[{{Primack} {et~al.}(1999){Primack}, {Bullock}, {Somerville}, \&
  {MacMinn}}]{1999APh....11...93P}
{Primack}, J.~R., {Bullock}, J.~S., {Somerville}, R.~S., \& {MacMinn}, D. 1999,
  Astroparticle Physics, 11, 93

\bibitem[{{Roustazadeh} \& {B{\"o}ttcher}(2010)}]{2010ApJ...717..468R}
{Roustazadeh}, P. \& {B{\"o}ttcher}, M. 2010, \apj, 717, 468

\bibitem[{{Roustazadeh} \& {B{\"o}ttcher}(2011)}]{2011ApJ...728..134R}
{Roustazadeh}, P. \& {B{\"o}ttcher}, M. 2011, \apj, 728, 134

\bibitem[{{Sbarufatti} {et~al.}(2006){Sbarufatti}, {Falomo}, {Treves}, \&
  {Kotilainen}}]{2006A&A...457...35S}
{Sbarufatti}, B., {Falomo}, R., {Treves}, A., \& {Kotilainen}, J. 2006, \aap,
  457, 35

\bibitem[{{Scarpa} {et~al.}(2000{\natexlab{a}}){Scarpa}, {Urry}, {Falomo},
  {Pesce}, \& {Treves}}]{2000ApJ...532..740S}
{Scarpa}, R., {Urry}, C.~M., {Falomo}, R., {Pesce}, J.~E., \& {Treves}, A.
  2000{\natexlab{a}}, \apj, 532, 740

\bibitem[{{Scarpa} {et~al.}(2000{\natexlab{b}}){Scarpa}, {Urry}, {Padovani},
  {Calzetti}, \& {O'Dowd}}]{2000ApJ...544..258S}
{Scarpa}, R., {Urry}, C.~M., {Padovani}, P., {Calzetti}, D., \& {O'Dowd}, M.
  2000{\natexlab{b}}, \apj, 544, 258

\bibitem[{{Schachter} {et~al.}(1993){Schachter}, {Stocke}, {Perlman}, {Elvis},
  {Remillard}, {Granados}, {Luu}, {Huchra}, {Humphreys}, {Urry}, \&
  {Wallin}}]{1993ApJ...412..541S}
{Schachter}, J.~F., {Stocke}, J.~T., {Perlman}, E., {et~al.} 1993, \apj, 412,
  541

\bibitem[{{Stecker} {et~al.}(1992){Stecker}, {de Jager}, \&
  {Salamon}}]{1992ApJ...390L..49S}
{Stecker}, F.~W., {de Jager}, O.~C., \& {Salamon}, M.~H. 1992, \apjl, 390, L49

\bibitem[{{Urry} {et~al.}(2000){Urry}, {Scarpa}, {O'Dowd}, {Falomo}, {Pesce},
  \& {Treves}}]{2000ApJ...532..816U}
{Urry}, C.~M., {Scarpa}, R., {O'Dowd}, M., {et~al.} 2000, \apj, 532, 816

\end{thebibliography}

\end{document}